\documentclass[aps,twocolumn,pra,10pt,longbibliography,superscriptaddress]{revtex4-2}
\usepackage{amssymb}
\usepackage{graphicx}
\usepackage{dcolumn}
\usepackage{bm}
\usepackage{amsmath}
\usepackage{subfigure}
\usepackage{float}
\usepackage{color}
\usepackage[colorlinks=true,citecolor=blue,urlcolor=blue]{hyperref}

\AtBeginEnvironment{thebibliography}{%
  \phantomsection\section{References}%
}

\usepackage{titlesec} 
\titleformat{\section}{\normalfont\large\bfseries}{}{0em}{} 
\titleformat{\subsection}{\normalfont\normalsize\bfseries}{}{0em}{} 

\begin{document}

\title{Observation of multiple time crystals in a driven-dissipative system with Rydberg gas}

\author{Yuechun Jiao}
\thanks{These authors contributed equally to this work.}
\affiliation{State Key Laboratory of Quantum Optics Technologies and Devices, Institute of Laser Spectroscopy, Shanxi University, Taiyuan 030006, China}
\affiliation{Collaborative Innovation Center of Extreme Optics, Shanxi University, Taiyuan 030006, China}

\author{Weilun Jiang}
\thanks{These authors contributed equally to this work.}
\affiliation{State Key Laboratory of Quantum Optics Technologies and Devices, Institute of Opto-Electronics, Shanxi University, Taiyuan 030006, China}
\affiliation{Collaborative Innovation Center of Extreme Optics, Shanxi University, Taiyuan 030006, China}

\author{Yu Zhang}
\affiliation{State Key Laboratory of Quantum Optics Technologies and Devices, Institute of Laser Spectroscopy, Shanxi University, Taiyuan 030006, China}

\author{Jingxu Bai}
\affiliation{State Key Laboratory of Quantum Optics Technologies and Devices, Institute of Laser Spectroscopy, Shanxi University, Taiyuan 030006, China}
\affiliation{Collaborative Innovation Center of Extreme Optics, Shanxi University, Taiyuan 030006, China}

\author{Yunhui He}
\affiliation{State Key Laboratory of Quantum Optics Technologies and Devices, Institute of Laser Spectroscopy, Shanxi University, Taiyuan 030006, China}

\author{Heng Shen}
\email{hengshen@sxu.edu.cn}
\affiliation{State Key Laboratory of Quantum Optics Technologies and Devices, Institute of Opto-Electronics, Shanxi University, Taiyuan 030006, China}
\affiliation{Collaborative Innovation Center of Extreme Optics, Shanxi University, Taiyuan 030006, China}

\author{Jianming Zhao}%
\email{zhaojm@sxu.edu.cn}
\affiliation{State Key Laboratory of Quantum Optics Technologies and Devices, Institute of Laser Spectroscopy, Shanxi University, Taiyuan 030006, China}
\affiliation{Collaborative Innovation Center of Extreme Optics, Shanxi University, Taiyuan 030006, China}

\author{Suotang Jia}%
\affiliation{State Key Laboratory of Quantum Optics Technologies and Devices, Institute of Laser Spectroscopy, Shanxi University, Taiyuan 030006, China}
\affiliation{Collaborative Innovation Center of Extreme Optics, Shanxi University, Taiyuan 030006, China}

\maketitle

\section{Abstract}
Time crystals, as temporal analogs of space crystals, manifest as stable and periodic behavior that breaks time translation symmetry. In an open quantum system, many-body interaction subjected to dissipation allows one to develop the time crystalline order in an unprecedented way, as refer to dissipative time crystals. Here we report the observation of multiple time crystals in the continuously driven-dissipative and strongly interacting Rydberg thermal gases, in which continuous time crystals, sub-harmonic time crystals and high-harmonic time crystals are observed in the same system by manipulating the Rydberg excitation. Our work provides new ways to explore the nonequilibrium phases of matter in open systems. Such time crystals with persistent oscillation rooted in emergent quantum correlations, may emerge as a ubiquitous tool in quantum metrology, for instance, continuous sensing and parameter estimation surpassing the standard quantum limit.

\section{Introduction}\label{sec1}
Crystal is a collection of atoms with the periodic arrangement in space, a celebrated example of spontaneous spatial translation symmetry breaking, ranging from salt and jewelry in daily life to bulk materials in the condensed matter experiment. We are accustomed to an inherent impression that time is on an equal footing with space, are thus curious whether time-translation symmetry can be spontaneously broken to form a time crystal. The original proposal of continuous time crystals (CTC) for isolated many-body systems in equilibrium~\cite{Wilczek2012} is prohibited by no-go theorems ~\cite{Bruno2013, Watanabe2015}. Instead, discrete time crystals (DTC) undergoing the spontaneous breaking of discrete time translation symmetry~\cite{Else2016,Yao2017,Zaletel2023} were discovered by applying a periodic external drive in an interacting many-body system under the ergodicity breaking. Such a closed system features a subharmonic response~\cite{Pal2018, Smits2018, Kyprianidis2021, Randall2021, Mi2022, Zhang2022}, and the associated time crystalline order is stabilized by disorder-induced many-body localization~\cite{Zhang2017,Choi2017,Randall2021} or prethermalization due to sufficiently high Floquet drive frequency~\cite{Kyprianidis2021,vu2023dissipative}.

An alternate strategy to stabilize the time crystalline order is dissipation~\cite{chitra2015dynamical,Gong2018,Buca2019,zhu2019dickea,gambetta2019DiscreteTimeCrystals,Booker2020,yao2020classicala, alaeian2021limit}, which is conventionally believed to destroy the order. Similar to reservoir engineering for preparing the desired quantum states, the discrete time crystalline order in a driven open Dicke model was realized by tailoring system-environment coupling appropriately~\cite{Kessler2021, Kongkhambut2021, Taheri2022}. Remarkably, following the pioneering theoretical works~\cite{Iemini2018,Buca2019}, a limit cycle phase without a recurring external force has also been confirmed recently in the platform of dissipative-driven Dicke model with ultra-cold atoms inside an optical cavity~\cite{Kongkhambut2022}. As the manifestation of CTC, a persistent oscillation robust against temporal perturbations was observed, taking the random phase between $0$ and $2\pi$ for different realizations~\cite{Kongkhambut2022,Chen2023}. In view of the foregoing, it is natural to raise a question, whether the interplay of strong interaction, dissipation and synchronization in the non-equilibrium open system could further enrich dynamic phases that spontaneously break time translation symmetry.

Here, we report the observation of a complex time crystalline order in the dissipative Rydberg gas at room temperature, exhibiting a series of spontaneous self-sustained oscillations of limit cycles under the time-independent external driving. Importantly, we perform the experiments with full control of the magnetic sub-states coupled by the light fields, the emergent limit cycles thus are unlikely to be caused by competing Rydberg states, as proposed in~\cite{wu2024}. Going beyond an inherent continuous time crystalline phase, we surprisingly find that sub-harmonic and high-order harmonic oscillations of limit cycles emerge successively in this open, quantum many-body system, regarded as sub-harmonic time crystals and high-harmonic time crystals, respectively. Such additional features of magnetic sub-state control are absent in previous works, including recent observations of CTC~\cite{wu2024}, a transient oscillation~\cite{Ding2024} and synchronized oscillation~\cite{Wadenpfuhl2023} in limit cycles with continuously driven Rydberg gas, leading to a furthering of knowledge on the phenomenon. Our work thus provides new understandings about a rich interplay between interaction, dissipation and synchronization in a quantum many-body system under continuous monitoring.

\begin{figure*}
\centering
\includegraphics[width=0.9\linewidth]{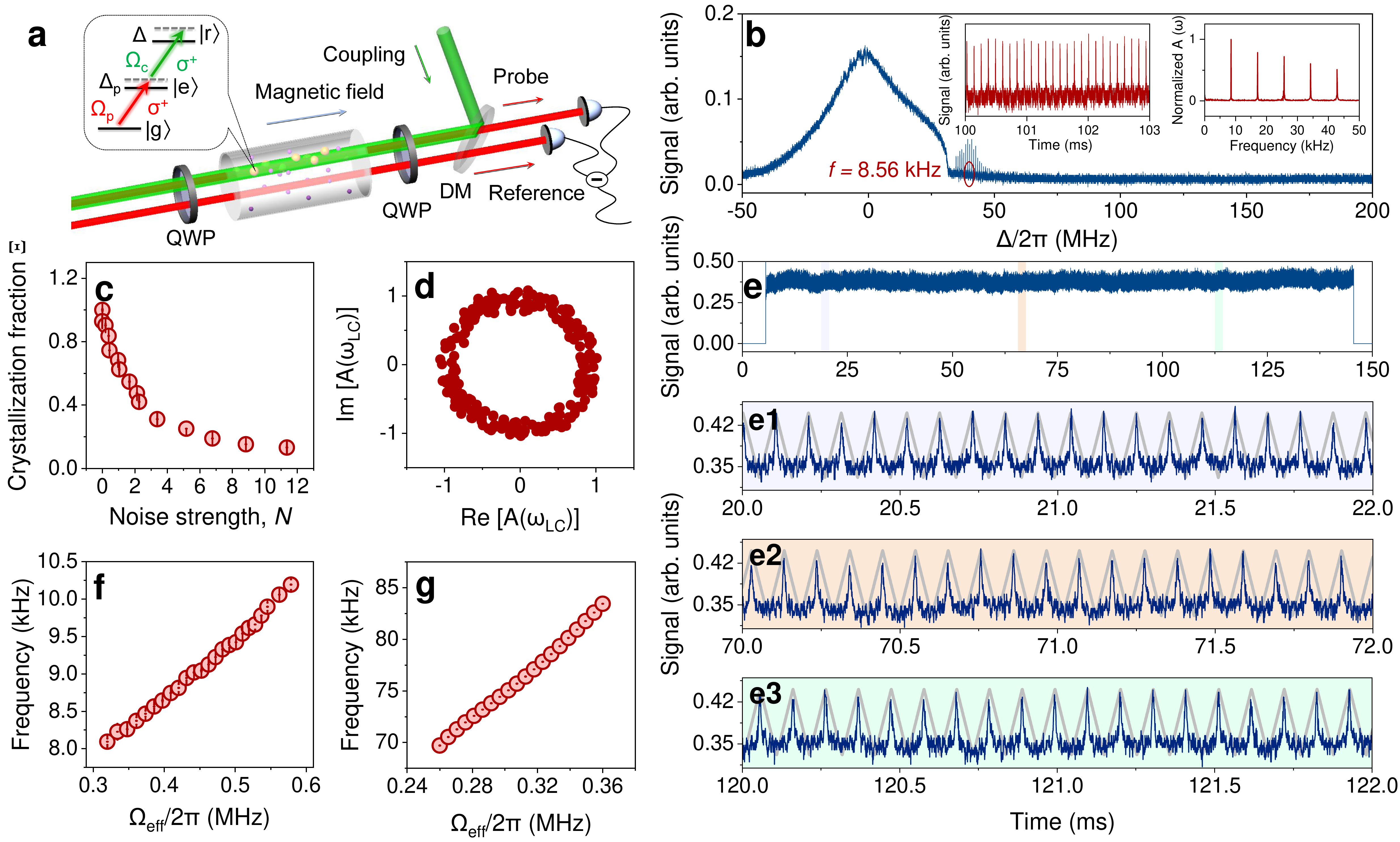}
\caption{\textbf{Emergence of continuous time crystals in a continuously driven-dissipative Rydberg system.} \textbf{a,} Schematic of the experimental setup and relevant energy level. QWP: quarter-wave plate, DM: dichroic mirror. \textbf{b,} Scanned transmission spectrum with $\Omega_p$ = 2$\pi \times$ 25.0~MHz and $\Omega_c$ = 2$\pi \times$ 2.3~MHz. Insets show the single-shot realization of the spontaneous self-sustained oscillation for $\Delta$ = 2$\pi \times$ 40~MHz and corresponding normalized single-sided amplitude spectra. \textbf{c,} Relative crystalline fraction $\Xi$ as a function of noise strength $N$. The error bars represent the standard deviation of two independent measurements. \textbf{d,} Distribution of the time phase in the limit cycle phase for 250 independent realizations. Both the data in c and d are taken at $\Omega_c$ = 2$\pi \times$ 2.3~MHz.  \textbf{e,} Demonstration of the single-shot realization of quench dynamics of CTC with an oscillation frequency of 9.615~kHz at $\Omega_c$ = 2$\pi \times$ 2.9~MHz.  \textbf{e1-e3}, Three segments of the 2-ms time window data trace, accompanied with a triangular reference waveform of frequency 9.615~kHz, corresponding to the gray marked time windows of 20-22~ms (\textbf{e1}), 70-72~ms (\textbf{e2}), and 120-122~ms (\textbf{e3}), respectively.  It is demonstrated that the phase shift of a single realization over 140~ms time window has a negligible effect. \textbf{f,} Measurements of the oscillation frequencies of CTC as a function of effective two-photon  Rabi frequency $\Omega_{\text{eff}}= \Omega_c \Omega_p /2\Delta_p$. The error bars are the fitting error for the fitting of single-sided amplitude spectra. In the experiment, the $\Omega_{\text{eff}}$ is varied by changing the coupling  Rabi frequency $\Omega_{c}$. \textbf{g,} Simulated oscillation frequencies of the time crystal according to the Hamiltonian in Eq.~\eqref{eq1}.
}
\label{Fig.1}
\end{figure*}

\section{Results}\label{sec2}
\subsection{Experimental setup}
We consider an ensemble of Rydberg atoms at room temperature. As illustrated in Fig.~\ref{Fig.1}\textbf{a}, two laser beams, referred as probe and coupling, excite the ground state $|g\rangle$ to the Rydberg state $|r\rangle$ via an intermediate state $|e\rangle$. This forms a ladder-type electromagnetically induced transparency (EIT) configuration, the associated strong interaction is given by the Hamiltonian $\hat{H}_I=\sum_{i<j} V_{i j} \hat{n}^r_i \hat{n}^r_j$ with the local Rydberg density $\hat{n}^r_i$ and the van der Waals interaction $ V_{i j}\propto -C_6/\left | \vec{r}_i-\vec{r}_j \right |^6$ between Rydberg atoms located at $\vec{r}_i$ and $\vec{r}_j$, with $C_6$ the dispersive coefficient ($C_6 <$0 for cesium 60$D_{5/2}$ atom used here).

To experimentally observe the non-equilibrium phases of matter in the thermal Rydberg gas, we use a resonant two-photon excitation scheme in a Cesium vapor cell. The counter-propagating probe (852 nm, $\Omega_p$) and control (509 nm, $\Omega_c$) fields set up the Rydberg EIT process (Fig.~\ref{Fig.1}\textbf{a}). A magnetic field $\mathbf{B}=11.8$~G along the propagation direction is applied to induce the large Zeeman splitting, thus ensuring the exact Zeeman level of the involving states can be specified by the polarization of the beams (See the Supplementary Note 2A). In this work, two laser beams with $\sigma^+$ circular polarization excite the transition of  $|g\rangle = |6S_{1/2}, F = 4, m_F = 4\rangle$ to $|r\rangle$ = $|60D_{5/2}, m_j = 5/2\rangle$ via the intermediate state $|e\rangle$ = $|6P_{3/2}, F^\prime = 5, m_F^\prime = 5\rangle$ with detuning of $\Delta_p$ = 2$\pi \times$70~MHz. A differential detection is used to suppress the laser intensity noise.

We stress that only single Rydberg state, $|60D_{5/2}, m_j = 5/2\rangle$, is involved in the driving scheme since we select the configuration of Rydberg transition via laser polarization, which distinguishes our finding sharply from the recent work~\cite{wu2024} where the appearance of oscillations is attributed to a competition of multiple Rydberg states. Scanned transmission spectra for the different polarizations are plotted in Supplementary Fig.~6.

\subsection{Observation of multiple time crystal phases}

On the experimental side, the high degree of controllability on Rydberg excitation facilitates the observation of time crystalline phases in the limit cycle. Here we perform a scanning EIT spectroscopy measurement by engineering the Rydberg excitation in terms of Rabi frequencies of the control field, $\Omega_c$. In experiments, we scan the coupling laser frequency at a typical scan rate of $2\pi \times$ 7~MHz/ms. Below a certain value of $\Omega_c = 2\pi \times $ 1.1~MHz, normal EIT spectra are observed, while the optical bistable phase emerges if exceeding it. By increasing $\Omega_c$ further, we observe oscillations of the transmission signal at the range of two-photon detuning $\Delta/2\pi \sim$ 35-45~MHz, and Fig.~\ref{Fig.1}\textbf{b} displays a typical transmission EIT spectrum exhibiting self-sustained oscillation at $\Omega_c$ = 2$\pi \times$ 2.3~MHz. Insets illustrate a typical realization of spontaneous self-sustained oscillation at the detuning $\Delta$ = 2$\pi \times$ 40~MHz. The associated normalized single-sided amplitude spectrum $A(\omega)$ is shown in the right panel. A narrow peak at $\omega_{LC}/2\pi$ = ~8.56 kHz with the full width at half maximum of 0.05~kHz (a time window of 16 ms) suggests a stable periodic oscillatory pattern (high frequency peaks display high harmonic components of oscillations).

As defined, the spontaneous breaking of continuous time translation symmetry and robustness against temporal perturbations constitute the essence of continuous time crystals. Regarding characterization of the robustness, we apply a white noise onto the probe beam, and use the relative crystalline fractions $\Xi$ to describe the robustness of the CTC on the noise strength $N$~\cite{Kongkhambut2022}. The relative crystalline fractions and the noise strength $N$ are defined by $\Xi$ = $\Sigma_{\omega =\omega_{LC}\pm\delta\omega}|A(\omega)|/\Sigma_{\omega}|A(\omega)|$ and $N=\Sigma_{\omega}|P_n(\omega)|/\Sigma_{\omega}|P_0(\omega)|-1$, respectively. $\delta\omega$ denotes the resolution of the single-sided amplitude, $P_n$ and $P_0$ are the single-sided amplitude spectra of the probe laser with and without white noise. Fig.~\ref{Fig.1}\textbf{c} shows the measured crystalline fraction as a function of the noise strength $N$. We also analyze the associated temporal dynamics, Fourier transformation and temporal correlation function of the oscillation spectra in the presence of noise, see Supplementary Fig.~8. Furthermore, the demonstration of long-term phase stability in the presence of noise \textit{N} = 2.1 is plotted in Supplementary Fig.~9. All the measurements in the presence of noise demonstrate that the oscillation exists over a wide parameter range, indicating the limit cycle phase is to some extent insensitive to the temporal perturbations.

In Fig.~\ref{Fig.1}\textbf{d}, we show the time phase of oscillations are randomly distributed over [0, 2$\pi$] for 250 independent realizations, indicative of spontaneous breaking of continuous time translation symmetry.  This observed random phase property is crucial to support the identification of spontaneous breaking of continuous time translation symmetry, and thus the time crystalline phase. Specifically, in order to rule out the experimental imperfections induced phase drift over time, Fig.~\ref{Fig.1}\textbf{e} displays a single-shot realization of quench dynamics exhibiting an oscillation frequency of 9.615~kHz at $\Omega_c$ = 2$\pi \times$ 2.9~MHz, where the probe field is suddenly switched on and held at a constant power. This 140 ms-time trace shows the good phase stability of a single realization on timescales spanning several independent realizations [Fig.~\ref{Fig.1}\textbf{e1-e3}]. Moreover, in such phase, we experimentally investigate the dependence of the oscillation frequency by varying the Rabi frequency of the control field. Given the fixed Rabi frequency of the probe field and bias magnetic field, the experimental result shows the oscillation frequency is linearly proportional to the coupling laser Rabi frequency, shown in Fig.~\ref{Fig.1}\textbf{f}, and its associated simulation results in Fig.~\ref{Fig.1}\textbf{g}.

\begin{figure*}[htbp]
\centering
\includegraphics[width=0.9\linewidth]{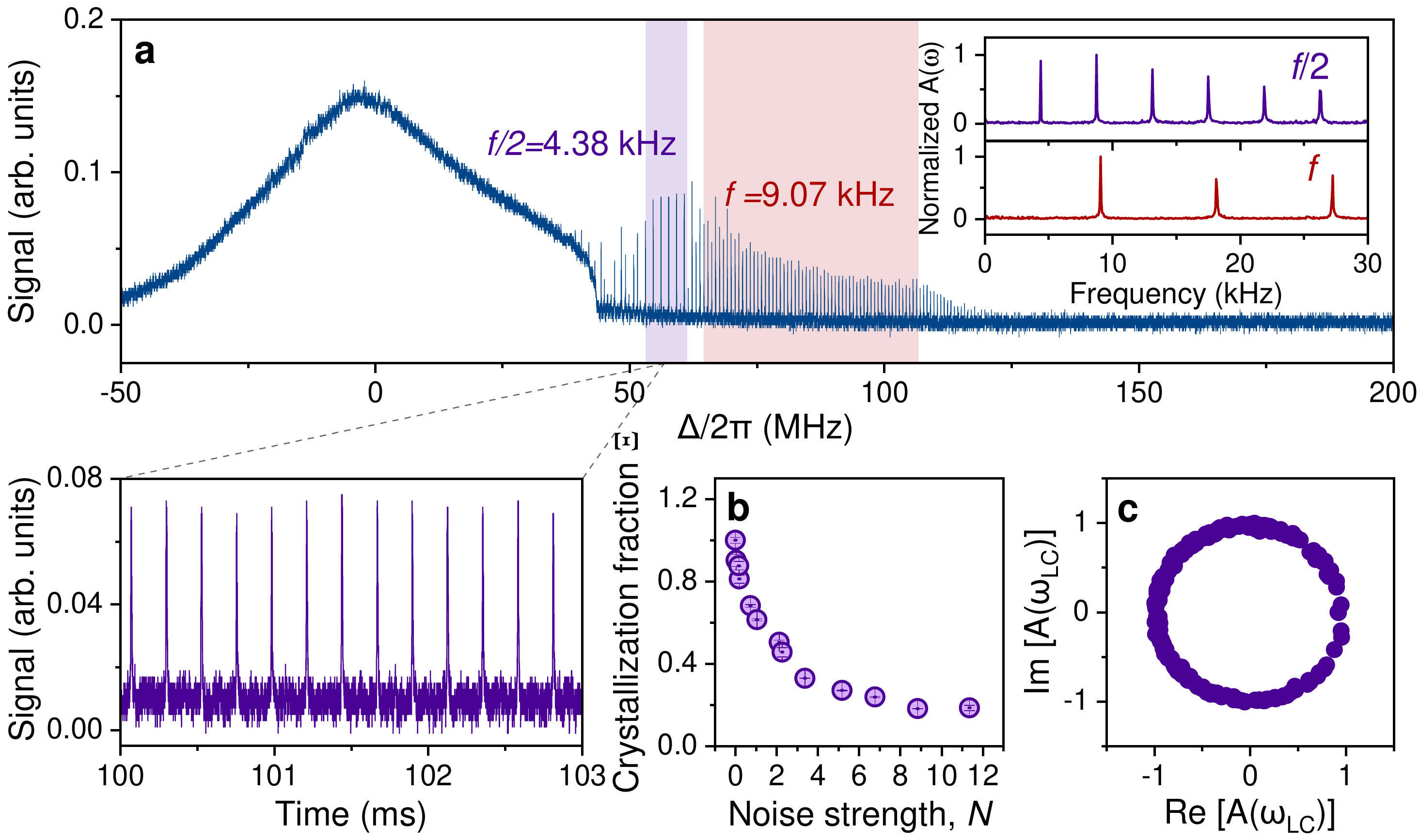}
\caption{\textbf{Observation of a self-sustained sub-harmonic time crystal.} \textbf{a,} Scanned transmission spectrum with $\Omega_c$ = 2$\pi \times$ 2.8~MHz. Inset: discrete Fourier transformation of the self-sustained oscillations in the parameter region marked in violet and red, respectively. \textbf{b,} Robustness of the observed subharmonic oscillatory pattern. The error bars represent the standard deviation of two independent measurements. \textbf{c,} Distribution of the time phase in the limit cycle phase.}
\label{Fig.2}
\end{figure*}

To explore the CTC phase further and probe the boundary of this phase, we slightly increase the coupling laser Rabi frequency $\Omega_c$ from 2$\pi \times$ 2.3~MHz to 2$\pi \times$ 2.8~MHz. Surprisingly, we find another persistent periodic oscillation at a lower frequency in the scanned transmission EIT spectrum (Fig.~\ref{Fig.2}\textbf{a}). By analyzing the frequency of two oscillation components for the detuning $\Delta$ = 2$\pi \times$ 57 MHz and 2$\pi \times$ 78 MHz, we find the new emergent oscillation has a frequency of 4.38 kHz (inset top panel in Fig.~\ref{Fig.2}\textbf{a}) that is half of the CTC oscillations, 9.07 kHz (inset bottom panel in Fig.~\ref{Fig.2}\textbf{a}). We also test the robustness of the limit cycle and the random distribution of its time phase to rule out the possibility of classical nonlinearity, shown in Figs.~\ref{Fig.2}\textbf{b} and \textbf{c}. All the observations demonstrate a sub-harmonic time crystal.

\begin{figure*}[htbp]
\centering
\includegraphics[width=0.9\textwidth]{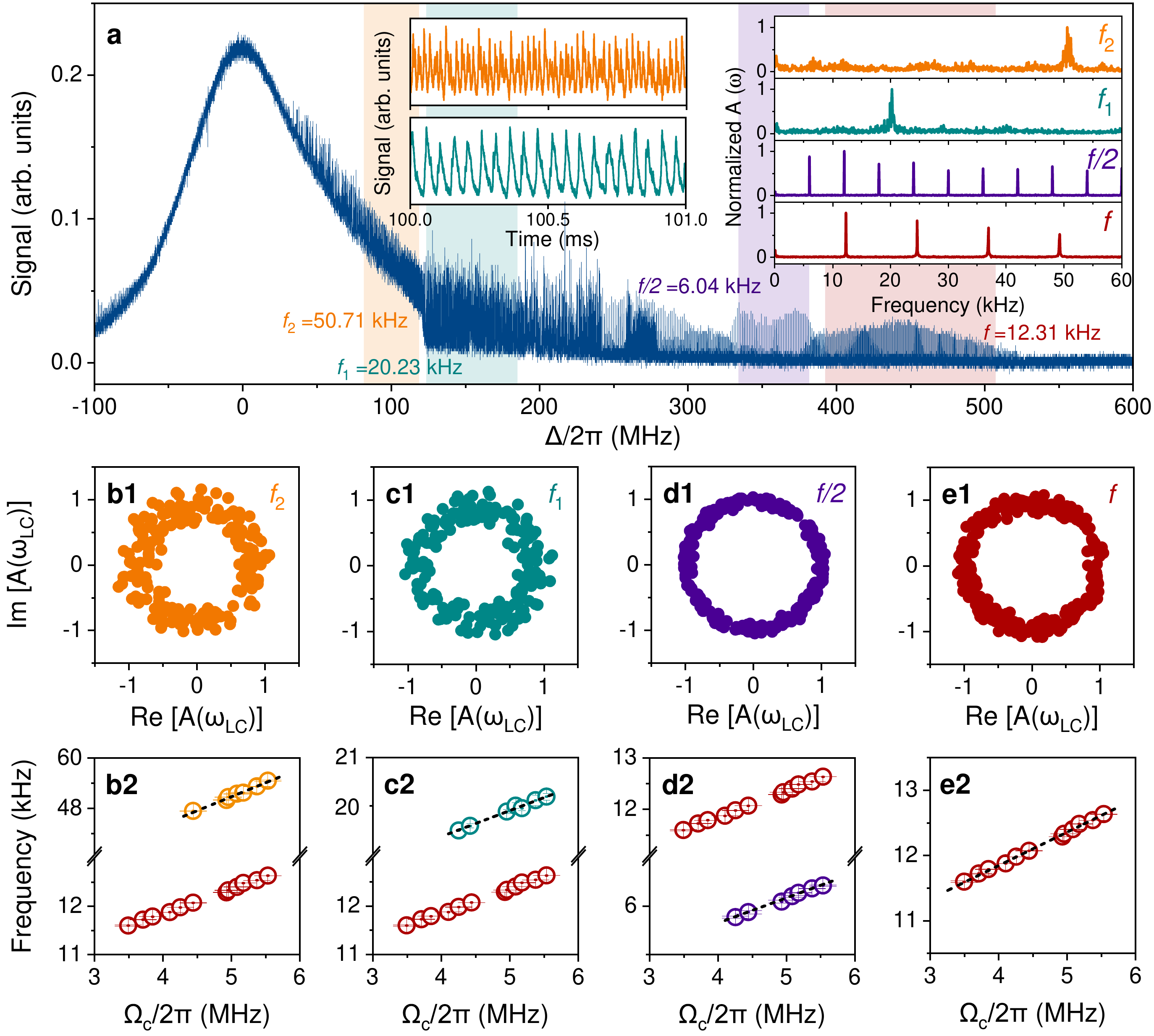}
\caption{\textbf{Observation of a self-sustained high-harmonic time crystal.} \textbf{a,} Scanned transmission spectrum with $\Omega_c$ = 2$\pi \times$ 5.0~MHz. Continuous time crystals, sub-harmonic and high-order time crystals (denoted as $f$, $f$/2, $f_1\approx 3f$/2 and $f_2 \approx 4f$) appear at different detuning, simultaneously. The corresponding oscillation frequencies are obtained by single-sided amplitude spectra. The inset shows the persistent high-order harmonic oscillations (left) and Fourier spectra of four oscillation components (right) at detunings of $\Delta/2\pi$ = 435 ($f$), 340 ($f/2$), 131 ($f_1$), and 110~MHz ($f_2$), respectively. \textbf{b1 - e1,} The distribution of the time phase in the limit cycle phase for high-harmonic, sub-harmonic, and continuous time crystals. \textbf{b2 - e2,} Measured oscillation frequency as a function of coupling Rabi frequency $\Omega_{c}$ for high-harmonic, sub-harmonic and continuous time crystals. The error bars are the fitting error for the fitting of single-sided amplitude spectra. Dashed lines illustrate the associated linear fit. The data of CTC frequency in \textbf{e2} are also plotted in \textbf{b2 - d2} (red circles) for easy comparison.
}
\label{Fig.3}
\end{figure*}

Experimentally, when lifting the coupling  Rabi frequency up to $2\pi \times$ 5.0~MHz, the scanned transmission EIT spectrum manifests emergent higher-order harmonics (labeled with $f_1$ and $f_2$), in addition to the CTC and sub-harmonic oscillations (labeled with $f$ and $f$/2), shown in Fig.~\ref{Fig.3}\textbf{a}. We extract their oscillation frequencies by discrete Fourier transform (DFT) with a time window of 16~ms, giving the values of 50.71~kHz, 20.23~kHz, 6.04~kHz, and 12.31~kHz, respectively, roughly corresponding to $f_2 \approx 4f$, $f_1\approx 3f$/2, $f$/2 and $f$, shown as the right inset. Subsequently, random distributions of their time phase with 250 repeated realizations are observed (Figs.~\ref{Fig.3}\textbf{b1-e1}), verifying the higher harmonics of limit cycle phases. Moreover, we demonstrate all oscillation frequencies increase linearly with the coupling laser Rabi frequency as we expected, shown in Figs.~\ref{Fig.3}\textbf{b2-e2}.

\subsection{Theoretical explanation}
To explain the observation of multiple time crystalline phases on the theoretical side, we consider a three-level model~\cite{Wadenpfuhl2023}. The associated Hamiltonian, $\hat H$, 
matching the experimental configuration in Fig.~\ref{Fig.1}, is written as,
\begin{equation}
    \begin{split}
	\hat{H}&=\sum_i\left[-\Delta_p  \hat n^e_i -(\Delta_c+\Delta_p)  \hat n^r_i + \frac{\Omega_p}{2} ( \hat \sigma^{ge}_i + \hat \sigma^{eg}_i)\right. \\&+\left. \frac{\Omega_c}{2} ( \hat \sigma^{er}_i + \hat \sigma^{re}_i) \right]+\sum_{i<j} V_{i j} \hat{n}^r_i \hat{n}^r_j,
    \end{split}
    \label{eq1}
\end{equation}
where $\hat n^\alpha_i = | \alpha \rangle_i \langle \alpha |_i$ is the population of level $\alpha$ of the $i$th atom, while $\hat \sigma^{\alpha \beta}_i = | \alpha \rangle_i  \langle \beta|_i$ with $\alpha, \beta = g,e,r$ represents the elements of atomic density operators associated with the ground state $|g\rangle$, the Rydberg state $|r\rangle$ and an intermediate state $|e\rangle$.

For a driven-dissipative system such as Rydberg gases, the system dynamics is  governed by the Lindblad master equation,
\begin{equation}
	\partial_t \hat \rho = \mathcal{L}(\hat \rho) = -i[\hat{H},\hat \rho]+ L^{ge}(\hat \rho) + L^{gr}(\hat \rho) + L^{er}(\hat \rho),  
    \label{eq2}
\end{equation}
where $L^{\alpha\beta}(\hat \rho)$ describes the dissipation channel between level $\alpha$ and $\beta$. 
In general, the competition between strong interaction-induced nonlinearity and dissipation can lead to complex mathematical solutions for the equations of motion (EOM). These solutions may exhibit riches of phenomena such as multistability, periodic oscillations, and other exotic non-equilibrium many-body phases. Examples include optical bistability~\cite{Carr2013}, self-organized criticality~\cite{Ding2020a}, and transitions toward synchronization~\cite{Wadenpfuhl2023}.

For the time-independent Hamiltonian, periodic oscillatory solutions imply the spontaneous breaking of continuous time-translation symmetry, also called CTC. Despite CTC in thermal Rydberg gases has been recently reported~\cite{wu2024}, a key outstanding question remains unresolved: the dependence of the oscillation frequency on system parameters.
To circumvent this limitation, we explicitly leverage the concept of boundary time crystal (BTC)~\cite{Iemini2018}. In fact, as stated in Ref.~\cite{Iemini2018}, the interacting Rydberg atomic system 
represents one of the most promising platforms for realizing BTC. In this framework, the interacting Rydberg atoms correspond to the boundary while the environment (the system bath) constitutes the bulk~\cite{russo2025quantum,wang2025boundary}. An oscillatory phase is induced by the interplay between driving and dissipation, with its characteristic frequency expected to scale linearly with the coupling Rabi frequency $\Omega_{c}$.
This analysis offers a theoretical explanation for the linear oscillatory frequency dependence observed in Fig.~\ref{Fig.1}\textbf{f}, in contrast with the previously observed inverse relationship to 
Rabi frequency in Ref.~\cite{wu2024}, in which the self-sustained oscillation originates from the competition of Rydberg excitations.

From a numerical perspective, it is impractical to directly solve the nonlinear master equation in Eq.~\eqref{eq2} for large systems. Here, we employ the mean-field (MF) approach, a standard treatment for thermal Rydberg gases~\cite{Carr2013, marcuzzi2014}, to address the strong nonlinear effect. Specifically, we develop a modified MF approach to include the high-order corrections, for instance, $\langle\hat{n}^r\rangle^2 \langle\hat{\sigma}^{gr}\rangle$,  $\langle\hat{n}^r\rangle^3 \langle\hat{\sigma}^{gr}\rangle$, stemming from the commutator between the Rydberg interaction term $\hat H_I$ and the density matrix elements. See Methods and Supplementary Note 1 for more details.

By combining the solutions of MF equation, we establish a theoretical framework explaining our observation of the multiple time crystalline. As demonstrated, the dynamics comprises three distinct phases: continuous time crystal, sub-harmonic time crystal, and high-harmonic time crystal.

(1) Continuous time crystal. Given the time-independent drive, an oscillatory solution of the observable $n^r$ is allowed in the parameter space. This originates from the nonlinearity ($\propto \hat{n}_i^r\hat{n}_j^r$) and is induced by the Rydberg interactions. Particularly, in such a three-level driven-dissipative system, three different phases with distinct time-evolution behavior can be identified, which are denoted as steady state (SS), bistable state (BS), oscillatory and bistable state (OBS), respectively. Concretely, in the SS phase, the Rydberg population $n^r$ approaches one stable solution, while this observable in the BS phase finally converges to two stable values. Strikingly, the dynamics of $n^r$ in the OBS phase exhibit the self-sustained oscillation for the upper branch solution, which is expressed as the co-existence of the oscillation and steady state. Consequently, we can find the self-sustained oscillation in this OBS phase, where the oscillation period is continuously varying. We regard it as the correspondence of the region with $f$ oscillatory frequency in Fig.~\ref{Fig.1}\textbf{b}, and in such phase we numerically find that the simulated frequency relation in Fig.~\ref{Fig.1}\textbf{g} is qualitatively in agreement with the experiments. Moreover, we also show the existence of Hopf bifurcation in the Supplementary Note 1B, which is used to interpret the emergence of the limit cycle. In combination with the robustness to perturbations and random phase distribution over [0, 2$\pi$] by different experimental repetitions given the phase stability of a single time trace as shown in Fig.~\ref {Fig.1}\textbf{e}, CTC can be identified.

(2) Sub-harmonic time crystal. Upon the occurrence of the CTC, when increasing the Rabi frequency of the coupling laser, we observe the subharmonic time crystal. We explain this phenomenon as a result of the periodic driving in the presence of bistability, as illustrated in Ref.~\cite{gambetta2019DiscreteTimeCrystals}. Here, the origin of this periodic driving is collisions-generated charged particles confined by a magnetic field, producing significant periodic electric fields that shift the resonance condition of the Rydberg excitation via the Stark effect in terms of the detuning $\Delta_c$. In fact, collisions between Rydberg atoms and ground-state atoms in the thermal Rydberg gases result in the generation of charged particles~\cite{weller2016chargeinduced,weller2019interplay}. The corresponding charge density nearly scales linearly with the Rydberg atom density. In short words, in this driven-dissipative system, CTC behaves as the self-sustained oscillation of the Rydberg atom density, leading to the periodically oscillating charge density and thus producing significant periodic electric fields. Importantly, the oscillation frequency of the generated electric fields is the same as that of the CTC, as a result of the linear proportional relationship between the charge density and the Rydberg atom density. Experimental verification can be found in Supplementary Fig.~14. Thus, the detuning $\Delta_c$ becomes time-dependent with the same oscillatory frequency as the oscillation frequency of the Rydberg atom density. In the nonlinear bistable regime, such an inherent oscillation behaves as a periodic drive and stimulates the system to perform the robust sub-harmonic response, forming a sub-harmonic time crystal phase. More details about the numerical simulation and additional experiments can be found in the Supplementary. 

(3) High-harmonic time crystal. At even a much stronger Rabi drive and lower detuning, the nonlinearity interaction dominates over other effects. Firstly, the average Rydberg excitation is expected to increase as $\Delta_c$ approaches the small detuning region, corresponding to $f_1$ and $f_2$ phase. Secondly, it has been predicted that the oscillation frequency scales with the Rabi frequency. Taken together, these observations suggest that a higher Rydberg population enhances the interactions, which in turn leads to an increase in the oscillatory frequency. To verify this in the numeric, we both reduce $\Delta_c$ and enlarge $\Omega_c$ to mimic the experimental conditions. In particular, we consider the average of trajectories from various initial conditions to describe the mixed state of the hot vapor. We find a larger oscillatory frequency as the parameters change, shown as Supplementary Fig.~5. The numerical results indicate that the emergence of higher frequency stable oscillations is related to the stronger interaction, which is comparable to the experimental findings. See Supplementary Note 1E for the details.

\section{Discussion}

We report the observation of a complex time crystalline order in an ensemble of thermal Rydberg gases under continuous driving. The observed self-sustained oscillations are recognized as continuous time crystals, sub-harmonic time crystals and high-harmonic time crystals. As an intrinsically out-of-equilibrium system, it provides a simple but versatile platform for exploring the exotic dynamic phases of open quantum systems~\cite{saffman2010quantuma, pizzi2019, yang2021, bakker2022drivendissipative}, such as macroscopic quantum synchronization effect~\cite{nadolny2023macroscopica,hajdusek2022seeding}. As suggested by recent work~\cite{choi2017arxiv}, time crystals observed here may be exploited in quantum metrology, such as continuous sensing surpassing the standard quantum limit~\cite{Cabot2024}, and time crystalline order as a frequency standard~\cite{Lyu2020}.

\section{Methods}

\subsection{Experimental details}
The experiment is conducted on a room-temperature Cs vapour cell (with a diameter of 2.5 cm and a length of 7.5 cm). The 852~nm and 509~nm lasers are delivered by Toptica DL pro and Precilasers, respectively. A high finesse ultralow expansion (ULE) cavity provides the feedback for the frequency stabilization of both lasers (FSR: 1.5 GHz, Finesse $1.5\times 10^5$). The tunable offset-lock frequency ensures the measurement of the limit cycle phase and robustness with fixed laser detuning. In addition, the 509~nm laser works in frequency-scanning mode for the measurement of the scanned EIT transmission spectrum. In this condition, the coupling laser frequency is calibrated by the cavity transmission spectrum. The $1/e^2$ beam waist of 852~nm and 509~nm are $\omega_p$ = 425~$\mu$m and $\omega_c$ = 450~$\mu$m, respectively. The homogeneous magnetic field $\mathbf{B}$ along with the probe direction is generated by a Helmholtz coil with 220~mm inner diameter. A probe and a reference beam pass through two identical acoustic optical modulators (AOMs) before entering the cell. When exploring the robustness against temporal perturbations of time crystals, we introduce white noise onto the probe beam by adding the white noise strength to modulate the RF amplitude of the AOM, rather than the reference beam.

\subsection{Theoretical details}

Here, we provide the details on the theoretical derivation. The dynamics of the Rydberg atom are well described by an effective three-level model. The Hamiltonian of the atomic system is defined in Eq.~\eqref{eq1}. As an open system, we consider the Master equation $\partial_t \hat \rho  = \mathcal{L}(\hat \rho) $ to describe the evolution of the system, with the dissipation,
\begin{equation}
	L^{\alpha \beta}(\hat O) = \Gamma^{\alpha \beta} \sum_i\left(\hat \sigma^{\alpha \beta}_i  \hat O \hat \sigma^{\beta \alpha}_i-\frac{1}{2}\left\{ \hat n^{\beta}_i , \hat O \right\}\right),
\end{equation}
where $\Gamma^{\alpha \beta}$ is the dissipation strength between level $\alpha$ and $\beta$. We consider a total of three decay channels in the simulation. The evolution for the average value of the observable $\hat O$ is, 
\begin{equation}
	\partial_t \langle \hat O \rangle=\langle  -i[\hat{H},\hat O]+ L^{ge}(\hat O) + L^{gr}(\hat O) + L^{er}(\hat O) \rangle.
\end{equation}
where $\langle \hat O \rangle$ is the average value of operator $\hat O$. The equation of motion for the observable on each site now reads,
{\footnotesize
\begin{equation}
	\begin{aligned}
	\partial_t\langle\hat{n}_i^g\rangle= & -\Omega_p \operatorname{Im}\langle\hat \sigma^{g e}_{i}\rangle+\Gamma^{g e} \langle \hat n_i^e\rangle+\Gamma^{g r} \langle \hat n_i^{r} \rangle\\
	\partial_t\langle\hat{n}_i^e\rangle= & +\Omega_p \operatorname{Im}\langle \hat \sigma^{g e}_i\rangle -\Omega_c \operatorname{Im}\langle\hat \sigma^{e r}_i\rangle-\Gamma^{g e} \langle \hat n^{e}_i\rangle+\Gamma^{e r} \langle \hat n^{r}_i \rangle\\
	\partial_t\langle\hat{n}_i^r\rangle= & +\Omega_c \operatorname{Im}\langle\hat \sigma^{e r}_i\rangle-\left(\Gamma^{g r}+\Gamma^{e r}\right) \langle \hat n^{r}_i \rangle\\
	\partial_t\langle\hat{\sigma}_i^{ge}\rangle= & -\frac{i}{2} \Omega_p\left(\langle \hat n^e_i\rangle-\langle \hat n^g_i \rangle \right)+\frac{i}{2} \Omega_c \langle \hat \sigma^{g r}_i \rangle \\
    & - i \Delta_p \langle \hat \sigma^{g e}_i \rangle-\frac{\Gamma^{g e}}{2} \langle \hat \sigma^{g e}_i \rangle \\
	\partial_t\langle\hat{\sigma}_i^{er}\rangle= & -\frac{i}{2} \Omega_c\left( \langle \hat n^r_i \rangle- \langle \hat n^e_i \rangle \right)-\frac{i}{2} \Omega_p \langle \hat \sigma^{g r}_i \rangle -i \Delta_c \langle \hat \sigma^{e r}_i \rangle \\
    &+ i \sum_j V_{ij} \langle \hat n^r_j \hat \sigma^{er}_i \rangle -\frac{\Gamma^{g e}+\Gamma^{e r}+\Gamma^{g r}}{2} \langle \hat \sigma^{e r}_i \rangle \\
	\partial_t\langle\hat{\sigma}_i^{gr}\rangle= & -\frac{i}{2} \Omega_p \langle \hat \sigma^{e r}_i \rangle+\frac{i}{2} \Omega_c \langle \hat \sigma^{g e}_i \rangle -\frac{\Gamma^{g r}+\Gamma^{e r}}{2} \langle \hat \sigma^{g r}_i \rangle \\
    &-i\left(\Delta_p+\Delta_c\right) \langle \hat \sigma^{g r}_i \rangle + i \sum_j V_{ij} \langle \hat n^r_j \hat \sigma^{gr}_i \rangle.
	\end{aligned}
	\label{eqS3}
\end{equation}
}
Since the Hamiltonian is translation invariant, we consider the solution with the translation symmetry, i.e., dropping the index $i$, $\langle \hat O_i \rangle$ = $\langle \hat O \rangle$. Moreover, it is worth noting that the commutator between the interaction term $\hat H_I$ and $\hat \sigma^{\alpha \beta} $ returns the term $i \sum_j V_{ij} \langle \hat n^{\beta}_j \hat \sigma^{\alpha \beta}_i \rangle$, resulting in the non-linearity and bring the difficulty to exactly solve. 

To simplify Eq.~\eqref{eqS3}, we employ the MF treatment.  The spirit of the MF is formularized as $\langle \hat \alpha \hat \beta \rangle \sim \langle \hat \alpha \rangle \langle \hat \beta \rangle$ at the lowest order~\cite{Carr2013, marcuzzi2014}. For example, it can be used in handling the term $\langle\hat{n}^r \hat{\sigma}^{gr} \rangle$ mentioned above. However, taking the strong interaction into account, the high-order term arising from the modification of the lowest order, such as $\langle\hat{n}^r\rangle^2 \langle\hat{\sigma}^{\alpha r}\rangle$,  $\langle\hat{n}^r\rangle^3 \langle\hat{\sigma}^{\alpha r}\rangle$. Concretely, $(\hat{n}^r)^x \hat{\sigma}^{\alpha r}$ is generated from the commutator between the $(\hat{n}^r)^{x-1} \hat{\sigma}^{\alpha r}$ and $\hat{\sigma}^{\alpha r}$. This high-order term will introduce more nonlinearity and further affect the solution, and may lead to novel physics. Here we expand the commutator term $\sum_j V_{ij} \langle \hat n^r_j \hat \sigma^{\alpha r}_i \rangle$  to the third order as an empirical choice,
\begin{equation}
V_1 \langle\hat{n}^r\rangle \langle\hat{\sigma}^{\alpha r}\rangle + V_2 \langle\hat{n}^r\rangle^2 \langle\hat{\sigma}^{\alpha r}\rangle+ V_3 \langle\hat{n}^r\rangle^3 \langle\hat{\sigma}^{\alpha r}\rangle,
\end{equation}
where $V_1, V_2, V_3$ are the corresponding coefficients for each order. After MF approximations, the EOM can be numerically solved by the Runge-Kutta method, and all time-dependent behaviors are thus able to be obtained. 

\section{Data availability} 
The data for Figs.~1-3 are provided in the source data file. The raw data are available from the corresponding author upon request.

\section{Code availability} The codes are available upon request from the corresponding author.
\vspace{15mm}

\bibliography{main}

\vspace{5mm}
\textbf{Acknowledgements} 
This work was supported by the National Key R\&D Program of China, No. 2020YFA0309400 (H.S.), NSFC of China, Nos. U2341211 (J.Z.), 12241408 (Y.J.), 62175136 (Y.J.), 12120101004 (J.Z.) and 12222409 (H.S.), and Innovation Program for Quantum Science and Technology, No. 2023ZD0300902 (Y.J.). H. S. acknowledges financial support from the Royal Society Newton International Fellowship Alumni follow-on funding (AL201024) of the United Kingdom and Fundamental Research Program of Shanxi Province (202203021223001).

\textbf{Author contributions} 
Y.J., H.S., and J.Z. conceived the idea. Y.J., Y.Z., and J.B. performed the experiments and collected the data. W.J. and Y.H. contributed to the theoretical simulation. Y.J., W.J., Y.H., and H.S. contributed to the data analysis.  Y.J., W.J., H.S., J.Z., and S.J. wrote the manuscript with contributions from all authors.

\textbf{Competing interests} The authors declare no competing interests.

\end{document}